\documentclass[a4paper,aps,showpacs,superscriptaddress]{revtex4}
\usepackage[T1]{fontenc}
\usepackage[latin9]{inputenc}
\usepackage{amsmath}
\usepackage{amssymb}
\usepackage{esint}

\makeatletter

\pdfpageheight\paperheight
\pdfpagewidth\paperwidth

\providecommand{\tabularnewline}{\\}

\@ifundefined{textcolor}{}
{%
 \definecolor{BLACK}{gray}{0}
 \definecolor{WHITE}{gray}{1}
 \definecolor{RED}{rgb}{1,0,0}
 \definecolor{GREEN}{rgb}{0,1,0}
 \definecolor{BLUE}{rgb}{0,0,1}
 \definecolor{CYAN}{cmyk}{1,0,0,0}
 \definecolor{MAGENTA}{cmyk}{0,1,0,0}
 \definecolor{YELLOW}{cmyk}{0,0,1,0}
}

\@ifundefined{date}{}{\date{}}

\usepackage{enumerate}\usepackage{subfigure}\usepackage{tabularx}

\newcommand{\be}{\begin{equation}}
\newcommand{\ee}{\end{equation}}
\newcommand{\ben}{\begin{eqnarray}}
\newcommand{\een}{\end{eqnarray}}
\newcommand{\bes}{\begin{subequations}}
\newcommand{\ees}{\end{subequations}}

\makeatother

\begin{document}

\title{Towards scaling cosmological solutions with full coupled Horndeski
Lagrangian: \\
 the KGB model}

\author{ A. R. Gomes}

\affiliation{Institut für Theoretische Physik, Universität Heidelberg, Philosophenweg
16, D-69120 Heidelberg, Germany.}

\affiliation{Departamento de F\'{i}sica, Instituto Federal do Maranhão, 65030-000
Sao Lu\'{i}s, MA, Brazil}

\author{Luca Amendola}

\affiliation{Institut für Theoretische Physik, Universität Heidelberg, Philosophenweg
16, D-69120 Heidelberg, Germany.}

\date{ }
\begin{abstract}
We study a general scalar field Lagrangian coupled with matter and
linear in $\Box\phi$ (also called KGB model). Within this class of
models, we find the most general form of the Lagrangian that allows
for cosmological scaling solutions, i.e. solutions where the ratio
of matter to field density and the equation of state remain constant.
Scaling solutions of this kind may help solving the coincidence problem
since in this case the presently observed ratio of matter to dark
energy does not depend on initial conditions, but rather on the theoretical
parameters. Extending previous results we find that it is impossible
to join in a single solution a matter era and the scaling attractor.
This is an additional step towards finding the most general scaling
Lagrangian within the Horndeski class, i.e. general scalar-tensor
models with second order equations of motion. 
\end{abstract}


\maketitle

\section{Introduction}

The search of suitable models based on scalar fields to explain the
accelerated expansion of the Universe \cite{Perlmutter_etal_1999,Riess_etal_1998}
is now more than ten years old. The main goal of this research has
been to find suitable solutions to the background and perturbation
equations of motion and to study their stability properties and their
degree of independence of the initial conditions. During the course
of this research the scalar field Lagrangian has been progressively
expanded by including terms coupled to gravity and terms that are
general functions of the kinetic energy. Recently some authors realized
that the most general scalar field Lagrangian that still produces
second order equations of motion is the so-called Horndeski Lagrangian
\cite{Horndeski:1974,Deffayet:2011gz,Kobayashi2}, a model that includes
four arbitrary functions of the scalar field and its kinetic energy.

An exhaustive study of the Horndeski model is very difficult due to
the number of free functions. It is therefore interesting to ask whether
one can find some general property without solving the equations of
motion. An important class of cosmological solutions that has been
studied for several models are the so-called scaling solutions, defined
by the property that the energy density of matter and scalar field
scale in the same way with time, so that their ratio remains constant.
A second condition that has also been often employed to simplify the
treatment is that the field equation of state remains constant. Scaling
solutions are particularly interesting because one can hope to employ
them to avoid the problem of the coincidence between the present matter
and dark energy densities, i.e. the fact that today the two density
fractions $\Omega_{m},\Omega_{\phi}$ are very similar. In fact, while
this coincidence occurs only today for a cosmological constant model
and for all the models in which matter and dark energy scale with
time in a different way, and therefore depends in a critical way on
the initial conditions, in scaling solutions the ``coincidence''
depends only the choice of parameters and, once established, can remain
true forever.

The prototypical case of scaling model is a simple uncoupled scalar
field with an exponential potential \cite{Copeland:1997et,Ferreira_Joyce_1998}.
However, this case can be immediately ruled out as a viable scaling
model since if pressureless matter is uncoupled then its equation
of state is zero and therefore any scaling component will also have
this equation of state, with the consequence that no acceleration
is possible during the scaling regime. The simplest way to solve this
problem and achieve scaling and acceleration is to couple the scalar
field and the matter component (or equivalently to couple field and
gravity) \cite{Wetterich_1995,amendola2000}. Several interesting
properties of this kind of scaling solutions have been studied in
the past, as for instance a similar coupling to neutrinos \cite{Amendola_Baldi_Wetterich_2008}
and the behavior of perturbations \cite{Amendola:2001rc}, and more
recently, with multiple dark matter models \cite{bbh,baldi}.

A powerful generalization of scaling models has been realized by Piazza
and Tsujikawa in Ref. \cite{pt} (see also \cite{ts}). They found
in fact that the most general Lagrangian without gravity coupling
that contains scaling solutions must have the form 
\begin{equation}
S=\int d^{4}x\sqrt{-g}\biggl[\frac{1}{2}R+K(\phi,X)\biggr]+S_{m}(\phi,\psi_{i},g_{\mu\nu})\label{action-1}
\end{equation}
with 
\begin{equation}
K(\phi,X)=Xg(Xe^{\lambda\phi}),
\end{equation}
where $X=-\frac{1}{2}\nabla_{\mu}\phi\nabla^{\mu}\phi$,\, $g$ an
arbitrary function and $\lambda$ a constant. $S_{m}$ is the action
for the matter fields, which also depends generally on the scalar
field $\phi$. The same form applies if the field has a constant coupling
to gravity. In Ref. \cite{Amendola:2006qi} this result has been extended
to variable couplings.

The scope of this paper is to perform another step in the direction
of extending this result to the entire Horndeski Lagrangian. We study
in fact a Lagrangian of type \cite{Kobayashi,Sawicki,Pujolas:2011he}
\begin{equation}
S=\int d^{4}x\sqrt{-g}\biggl[\frac{1}{2}R+K(\phi,X)-G_{3}(\phi,X)\nabla_{\mu}\nabla^{\mu}\phi\biggr]+S_{m}(\phi,\psi_{i},g_{\mu\nu})\label{action-2}
\end{equation}
denoted KGB model in \cite{Sawicki}. The new term containing $G_{3}$
produces new second order terms in the equation of motion. As we will
see, the addition of the term linear in $\Box\phi\equiv\nabla_{\mu}\nabla^{\mu}\phi$
introduces several new features and enlarges considerably the class
of models that allow for accelerated scaling solutions. However, we
will also find that the properties of the scaling solutions are essentially
unchanged. The scaling expansion law in fact does not depend on the
new term in $G_{3}$.

Before concluding we will also show that it is not possible to reach
the scaling solution after a standard matter dominated era.


\section{Horndeski Lagrangian and equations of motion}

As anticipated, we consider an action consisting in 
\begin{equation}
S=\int d^{4}x\sqrt{-g}\biggl[\frac{1}{2}R+K(\phi,X)-G_{3}(\phi,X)\nabla_{\mu}\nabla^{\mu}\phi\biggr]+S_{m}(\phi,\psi_{i},g_{\mu\nu})\label{action}
\end{equation}
where $\phi$ is a scalar field and $X=-\frac{1}{2}\nabla_{\mu}\phi\nabla^{\mu}\phi$.
This action is part of the more general Horndeski Lagrangian and as
such gives rise to second order equations of motion. We consider that
there is only one type of matter of energy density $\rho_{m}=-T_{0}^{0}$,
in the Einstein frame, where the energy-momentum tensor is defined
by 
\begin{equation}
T_{\mu\nu}=-\frac{2}{\sqrt{-g}}\frac{\delta S_{m}}{\delta g^{\mu\nu}}.
\end{equation}
In this frame matter is directly coupled to the scalar field through
the function $Q(\phi)$, where 
\begin{equation}
Q=-\frac{1}{\rho_{m}\sqrt{-g}}\frac{\delta S_{m}}{\delta\phi}.\label{Qdef}
\end{equation}
A comparative analysis between this and the formulation in the Jordan
frame is presented in Sec. \ref{sec:So-far-we} of this work. In particular,
we show that scaling solutions in a frame remain scaling in the other
as well.

Eq. (\ref{action}) has the form $S=S_{(E-H)}+S_{2}+S_{3}+S_{m}$,
where $S_{E-H}$ is the Einstein-Hilbert action, $S_{2}$ depends
on $K(\phi,X)$ and $S_{3}$ depends on $G_{3}(\phi,X)\nabla_{\mu}\nabla^{\mu}\phi$.
Integrating $S_{3}$ by parts we can arrive at an equivalent action
\cite{Sawicki}: 
\begin{eqnarray}
S_{3} & = & -\int d^{4}x\sqrt{-g}G_{3}(\phi,X)\nabla_{\mu}\phi\nabla^{\mu}\phi\nonumber \\
 & = & \int d^{4}x\sqrt{-g}[G_{3,\phi}\nabla_{\mu}\phi+G_{3,X}\nabla_{\mu}X]\nabla^{\mu}\phi.
\end{eqnarray}
In this work we are using $G_{3,\phi}=\partial G_{3}/\partial\phi$,
$G_{3,X}=\partial G_{3}/\partial X$, $G_{3,\phi X}=\partial^{2}G_{3}/(\partial\phi\partial X)$
and similar simplifying notations for other partial derivatives of
$G_{3}(\phi,X)$ and $K(\phi,X)$. The former expression for the action
shows that the Lagrangian density 
\begin{equation}
p=K+G_{3,\phi}\nabla_{\mu}\phi\nabla^{\mu}\phi+G_{3,X}\nabla_{\mu}X\nabla^{\mu}\phi\label{pdef}
\end{equation}
is equivalent to the original Lagrangian density 
\begin{equation}
\mathcal{L}=K(\phi,X)-G_{3}(\phi,X)\Box\phi.\label{originallagrangian}
\end{equation}
Therefore $p$ can be at most linear in $\Box\phi$, a condition we
will use further below. We consider a FLRW flat metric with $ds^{2}=-dt^{2}+\mathcal{A}^{2}(t)d\textbf{x}^{2}$,
where $\mathcal{A}(t)$ is the scale factor. In this case we have
$X=\dot{\phi}^{2}/2$, $\dot{X}=\dot{\phi}\ddot{\phi}$ and $\nabla_{\mu}X\nabla^{\mu}\phi=-2X\ddot{\phi}$,
where dot means derivative with respect to the cosmic time $t$. Then
we can write Eq. (\ref{pdef}) as 
\begin{equation}
p=K-2X(G_{3,\phi}+\ddot{\phi}G_{3,X}).\label{press}
\end{equation}
The energy-momentum tensor of the scalar field is defined as 
\begin{equation}
T_{\mu\nu}^{(\phi)}=-\frac{2}{\sqrt{-g}}\frac{\delta(S_{2}+S_{3})}{\delta g^{\mu\nu}}.
\end{equation}
The pressure $p_{\phi}=T_{1}^{1}=T_{2}^{2}=T_{3}^{3}$ is identified
with the $p$ found previously. The energy density of the scalar field
$\rho_{\phi}=-T_{0}^{0}$ is found to be 
\begin{equation}
\rho_{\phi}=2XK_{X}-K-2XG_{3,\phi}+6X\dot{\phi}HG_{3,X}.\label{rho_KG3}
\end{equation}
Varying the action $S$ with respect to $g^{\mu\nu}$ gives 
\begin{equation}
H^{2}=\frac{1}{3}(\rho_{\phi}+\rho_{m})\label{feq}
\end{equation}
and 
\begin{equation}
3H^{2}+2\dot{H}=-p-p_{m},\label{feq2}
\end{equation}
with $H=\dot{\mathcal{A}}/\mathcal{A}$. We define also $\rho_{t}=\rho_{\phi}+\rho_{m}$
and $p_{t}=p_{\phi}+p_{m}$. Defining 
\begin{equation}
\Omega_{\phi}=\frac{\rho_{\phi}}{3H^{2}},\,\,\,\Omega_{m}=\frac{\rho_{m}}{3H^{2}}\label{Omegaphi}
\end{equation}
we can rewrite the Friedman equation (Eq. (\ref{feq})) as 
\begin{equation}
\Omega_{\phi}+\Omega_{m}=1.\label{sumOmega}
\end{equation}
Now we introduce $d/dt=Hd/dN$. Then the equation of motion for the
scalar field $\phi$ and matter are \cite{horn} 
\begin{align}
\frac{d\rho_{\phi}}{dN}+3(1+w_{\phi})\rho_{\phi} & =-\rho_{m}Q\frac{d\phi}{dN}\label{rhophi2}\\
\frac{d\rho_{m}}{dN}+3(1+w_{m})\rho_{m} & =\rho_{m}Q\frac{d\phi}{dN}.\label{rhom2}
\end{align}
where $w_{\phi}=p/\rho_{\phi}$. A useful relation is also 
\begin{equation}
\frac{\dot{H}}{H^{2}}=-\frac{3}{2}(1+w_{eff}).\label{dotHH2}
\end{equation}
where $w_{eff}=\Omega_{m}w_{m}+\Omega_{\phi}w_{\phi}$.


\section{Scaling Solutions}

The condition $\Omega_{\phi}/\Omega_{m}$ constant define scaling
solutions. This is equivalent to $\rho_{\phi}/\rho_{m}$ constant,
or to 
\begin{equation}
\frac{d\log\rho_{\phi}}{dN}=\frac{d\log\rho_{m}}{dN}\label{scaling}
\end{equation}
Also, from Eq. (\ref{sumOmega}) we get that $\Omega_{\phi}$ is a
constant. We also assume that for asymptotic scaling solutions the
equation of state parameter $w_{\phi}$ is a constant \cite{ts}.
Subtracting both Eqs. (\ref{rhophi2}) and (\ref{rhom2}) and using
Eq. (\ref{scaling}) we get 
\begin{equation}
\frac{d\phi}{dN}=\frac{3\Omega_{\phi}}{Q}(w_{m}-w_{\phi})\propto\frac{1}{Q}.\label{dphidN}
\end{equation}
Back to Eqs. (\ref{rhophi2}) and (\ref{rhom2}) we get 
\begin{equation}
\frac{d\log\rho_{\phi}}{dN}=\frac{d\log\rho_{m}}{dN}=-3(1+w_{eff}),
\end{equation}
Now, from $w_{\phi}$ constant, we have 
\begin{equation}
\frac{d\log p}{dN}=-3(1+w_{eff})
\end{equation}
We want to find a covariant master equation for $p=p(X,\Box{\phi},\phi)$.
The former equation gives 
\begin{equation}
\frac{\partial\log p}{\partial\log X}\frac{d\log X}{dN}+\frac{\partial\log p}{\partial\log\Box\phi}\frac{d\log\Box\phi}{dN}+\frac{\partial\log p}{\partial\phi}\frac{d\phi}{dN}=-3(1+w_{eff}).\label{master}
\end{equation}
We need the partial derivatives ${d\log X}/{dN}$ and ${d\log\Box\phi}/{dN}$,
that are obtained as follows:

\subsection{${d\log X}/{dN}$}

From the definition of $X$ and Eq. (\ref{dphidN}) we have 
\begin{equation}
X=\frac{1}{2}\dot{\phi}^{2}=\frac{H^{2}}{2}\biggl(\frac{d\phi}{dN}\biggr)^{2}\propto\frac{H^{2}}{Q^{2}}\propto\frac{p}{Q^{2}},
\end{equation}
and then 
\begin{eqnarray}
\frac{d\log X}{dN} & = & \frac{d\log p}{dN}-2\frac{d\log Q}{dN}\nonumber \\
 & = & -3(1+w_{eff})-\frac{2}{Q}\frac{dQ}{dN}
\end{eqnarray}

\subsection{${d\log\Box\phi}/{dN}$}

We start with 
\begin{equation}
\Box\phi=-3H\dot{\phi}-\ddot{\phi}\label{Boxphi0}
\end{equation}
Now, from Eq. (\ref{dphidN}) this can be rewritten as 
\begin{equation}
\Box\phi=-\frac{3}{2}\frac{w_{m}-w_{\phi}}{w_{\phi}}(1-w_{eff})\frac{p}{Q}\biggl[1-\frac{2}{\lambda}\frac{1+w_{eff}}{1-w_{eff}}\frac{1}{Q^{2}}\frac{dQ}{d\phi}\biggr],\label{Boxphi}
\end{equation}
with 
\begin{equation}
\lambda=\frac{1+w_{eff}}{\Omega_{\phi}(w_{m}-w_{\phi})}.\label{lambda}
\end{equation}
So far we put no restirctions on the coupling function $Q$. However
we find that the analysis is very simplified if we assume 
\begin{equation}
\frac{1}{Q^{2}}\frac{dQ}{d\phi}=const.\label{eqQ}
\end{equation}
This restricts the coupling to be 
\begin{equation}
Q(\phi)=\frac{1}{c_{1}\phi+c_{2}},\label{Qsol}
\end{equation}
with $c_{1}$, $c_{2}$ constants. Later on, however, we will specialize
to the case of constant $Q$. From Eq. (\ref{Boxphi}) we have then
\begin{eqnarray}
\frac{d\log\Box\phi}{dN} & = & \frac{d\log p}{dN}-\frac{d\log Q}{dN}\nonumber \\
 & = & -3(1+w_{eff})-\frac{1}{Q}\frac{dQ}{dN}.
\end{eqnarray}
Finally, Eq. (\ref{master}) becomes 
\begin{equation}
\biggl(1+\frac{2}{\lambda Q^{2}}\frac{dQ}{d\phi}\biggr)\frac{\partial\log p}{\partial\log X}+\biggl(1+\frac{1}{\lambda Q^{2}}\frac{dQ}{d\phi}\biggr)\frac{\partial\log p}{\partial\log\Box\phi}-\frac{1}{\lambda Q}\frac{\partial\log p}{\partial\phi}=1.\label{master2}
\end{equation}
As expected, the master equation reduces to the one obtained in Ref.
\cite{Amendola:2006qi} when $G_{3}(\phi,X)=0$.


\section{Solutions for the master equation}

Here, after a convenient \textit{Ansatz}, we derive the general solution
for the master equation Eq. (\ref{master2}). Remember, however, that
there are restrictions in the form of $Q(\phi)$, given by Eq. (\ref{Qsol}),
that will be taken into account in due course. We start with Eq. (\ref{master2})
rewritten as 
\begin{equation}
\biggl(1+\frac{2}{Q}\frac{dQ}{d\psi}\biggr)\frac{\partial\log p}{\partial\log X}+\biggl(1+\frac{1}{Q}\frac{dQ}{d\psi}\biggr)\frac{\partial\log p}{\partial\log\Box\phi}-\frac{\partial\log p}{\partial\psi}=1,\label{master3}
\end{equation}
where 
\begin{equation}
\psi=\int_{\phi}du[\lambda Q(u)].\label{psi-def}
\end{equation}
Set 
\begin{equation}
p=XQ^{2}(\phi)\tilde{g}(X,\Box\phi,\phi).\label{p-tildeg-Q}
\end{equation}
where $\tilde{g}$ is an arbitray function of its argument. Then for
$\tilde{g}\neq0$ we obtain 
\begin{equation}
\biggl(1+\frac{2}{Q}\frac{dQ}{d\psi}\biggr)X\frac{\partial\tilde{g}}{\partial X}+\biggl(1+\frac{1}{Q}\frac{dQ}{d\psi}\biggr)\Box\phi\frac{\partial\tilde{g}}{\partial\Box\phi}-\frac{\partial\tilde{g}}{\partial\psi}=0.\label{masterQ}
\end{equation}
where by (\ref{eqQ}) the term $\frac{2}{Q}\frac{dQ}{d\psi}$ is a
constant. This partial differential equation is linear in $\tilde{g}$.
Then the method of separation of variables is justifiable, and the
general solution must be of the form 
\begin{equation}
\tilde{g}=g_{a}(h_{a})+g(h_{b})+g_{c}(h_{c})+g_{d}(h_{d}),\label{ga-gd-Q}
\end{equation}
where $g_{a},g,g_{c},g_{d}$ are arbitrary functions and 
\begin{eqnarray}
h_{a}(X,\Box\phi,\psi) & = & f_{1a}(X)f_{2a}(\Box\phi)f_{3a}(\psi),\label{haQ}\\
h_{b}(X,\psi) & = & f_{1b}(X)f_{3b}(\psi),\label{hbQ}\\
h_{c}(X,\Box\phi) & = & f_{1c}(X)f_{2c}(\Box\phi)\label{hcQ}\\
h_{d}(\Box\phi,\psi) & = & f_{2d}(\Box\phi)f_{3d}(\psi).\label{hdQ}
\end{eqnarray}
In the following we will consider separately the four functions.

\subsection{$g_{a}(h_{a})$}

Eq. (\ref{masterQ}) gives 
\begin{equation}
\frac{dg_{a}}{dh_{a}}\biggl[\biggl(1+\frac{2}{Q}\frac{dQ}{d\psi}\biggr)\frac{1}{f_{1a}}\frac{df_{1a}}{d\log X}+\biggl(1+\frac{1}{Q}\frac{dQ}{d\psi}\biggr)\frac{1}{f_{2a}}\frac{d\log f_{2a}}{d\log\Box\phi}-\frac{1}{f_{3a}}\frac{\partial f_{3a}}{\partial\psi}\biggr]=0.\label{eq:masterQa}
\end{equation}
By separation of variables we find that we can take $\log f_{1a}=\alpha\log X$
and $\log f_{2a}=\beta\log\Box\phi$. Then Eq. (\ref{eq:masterQa})
gives $f_{3a}=e^{(\alpha+\beta)\psi}Q^{2\alpha+\beta}$. Then Eq.
(\ref{haQ}) gives 
\begin{equation}
h_{a}=\biggl[X(\Box\phi)^{\beta/\alpha}e^{(1+\beta/\alpha)\psi}Q^{2+\beta/\alpha}\biggr]^{\alpha}
\end{equation}
and 
\begin{equation}
g_{a}(h_{a})=g_{a}\biggl(X(\Box\phi)^{\beta/\alpha}e^{(1+\beta/\alpha)\psi}Q^{2+\beta/\alpha}\biggr),
\end{equation}
where the exponent $\alpha$ from $h_{a}$ was absorbed since $g_{a}$
is a general function. Similar procedure is done in obtaining the
other functions $g,g_{c},g_{d}$. As $p$ is expected to be at most
linear in $\Box\phi$ (see eq. (\ref{originallagrangian})), we choose
$\beta/\alpha=1$ in the former equation. This leads to 
\begin{equation}
g_{a}(h_{a})=g_{a}\biggl(X(\Box\phi)e^{2\psi}Q^{3}(\phi)\biggr).
\end{equation}

\subsection{$g(h_{b})$}

Eq. (\ref{masterQ}) gives 
\begin{equation}
\frac{dg}{dh_{b}}\biggl[\biggl(1+\frac{2}{Q}\frac{dQ}{d\psi}\biggr)\frac{d\log f_{1b}}{d\log X}-\frac{1}{f_{3b}}\frac{\partial f_{3b}}{\partial\psi}\biggr]=0,
\end{equation}
which gives $f_{1b}=X^{\alpha}$ and $f_{3b}=e^{\alpha\psi}Q^{2\alpha}$.
Then Eq. (\ref{hbQ}) gives 
\begin{equation}
h_{b}=(XQ^{2}(\phi)e^{\psi})^{\alpha}
\end{equation}
and therefore 
\begin{equation}
g(h_{b})=g(XQ^{2}(\phi)e^{\psi}).
\end{equation}

\subsection{$g_{c}(h_{c})$}

Eq. (\ref{masterQ}) gives 
\begin{equation}
\frac{dg_{c}}{dh_{c}}\biggl[\biggl(1+\frac{2}{Q}\frac{dQ}{d\psi}\biggr)\frac{d\log f_{1c}}{d\log X}+\biggl(1+\frac{1}{Q}\frac{dQ}{d\psi}\biggr)\frac{d\log f_{2c}}{d\log\Box\phi}\biggr]=0.\label{master-gc-Q}
\end{equation}
If we consider now $f_{1c}=X^{\alpha}$ and $f_{2c}=(\Box\phi)^{\beta}$,
the only solution compatible with Eq. (\ref{master-gc-Q}) and the
requirement of no explicit dependence of $g_{c}$ on $\phi$ is $\alpha=\beta=0$.
This shows that $g_{c}(h_{c})=0$ for $Q(\phi)\neq0$. This must be
compared with the functional dependence $g_{c}\biggl({\Box\phi}/{X}\biggr)$
obtained for constant $Q$. In that case the requirement of linearity
of $p$ with $\Box\phi$ results in a trivial constant to be added
to $G_{3}$. Then, we conclude that for all $Q(\phi)$ obeying the
master equation there is no influence in the equations of motion.
In this way we can discard the $g_{c}(h_{c})$ term for the Lagrangian
with scaling solutions.

\subsection{$g_{d}(h_{d})$}

Eq. (\ref{masterQ}) gives 
\begin{equation}
\frac{dg_{d}}{dh_{d}}\biggl[\biggl(1+\frac{1}{Q}\frac{dQ}{d\psi}\biggr)\frac{d\log f_{2d}}{d\log\Box\phi}-\frac{1}{f_{3d}}\frac{\partial f_{3a}}{\partial\psi}\biggr]=0.\label{master-gd-Q}
\end{equation}
which gives $f_{2d}=(\Box\phi)^{\alpha}$ and $f_{3d}=Q^{\alpha}e^{\alpha\psi}$.
Then Eq. (\ref{hdQ}) gives 
\begin{equation}
h_{d}=\biggl((\Box\phi)e^{\psi}Q\biggr)^{\alpha}
\end{equation}
and therefore 
\begin{equation}
g_{d}(h_{d})=g_{d}\biggl((\Box\phi)Q(\phi)e^{\psi}\biggr).
\end{equation}
Finally, from the former results and Eqs. (\ref{p-tildeg-Q}) and
(\ref{ga-gd-Q}) we obtain 
\begin{equation}
p(X,\Box\phi,\phi)=XQ^{2}(\phi)\biggl[g_{a}\biggl(XQ^{3}(\phi)(\Box\phi)e^{2\psi}\biggr)+g(XQ^{2}(\phi)e^{\psi})+g_{d}\biggl(Q(\phi)(\Box\phi)e^{\psi}\biggr)\biggr].\label{pQ0}
\end{equation}
Note that $g_{a}=g_{d}=0$ gives 
\begin{equation}
p=XQ^{2}(\phi)g(XQ^{2}(\phi)e^{\psi}),\label{pQ0-prd}
\end{equation}
which is the known result from Ref. \cite{Amendola:2006qi}. The restriction
that, in general, $p(X,\Box\phi,\phi)$ must be at most linear in
$\Box\phi$ gives 
\begin{equation}
p(X,\Box\phi,\phi)=XQ^{2}(\phi)\biggl[-a\biggl(XQ^{3}(\phi)(\Box\phi)e^{2\psi}\biggr)+g(XQ^{2}(\phi)e^{\lambda Q\phi})-r\biggl(Q(\phi)(\Box\phi)e^{\psi}\biggr)\biggr],
\end{equation}
where $a,r$ are arbitrary constants and $g$ remains a general function.
We know that $p$ is equivalent to the initial Lagrangian density,
$\mathcal{L}=K(\phi,X)-G_{3}(\phi,X)\Box\phi$. If we take now $p=\mathcal{L}$,
we get 
\begin{equation}
K(\phi,X)=XQ^{2}(\phi)g(XQ^{2}(\phi)e^{\psi})\label{K}
\end{equation}
and 
\begin{equation}
G_{3}(\phi,X)=XQ^{2}(\phi)\biggl[a\biggl(XQ^{3}(\phi)e^{2\psi}\biggr)+r\biggl(Q(\phi)e^{\psi}\biggr)\biggr],
\end{equation}
We can rewrite the former equation as 
\begin{equation}
G_{3}(\phi,X)=aX^{2}Q^{5}(\phi)e^{2\psi}+rXQ^{3}(\phi)e^{\psi}.\label{G3}
\end{equation}
and the Lagrangian as 
\begin{equation}
\mathcal{L}(X,\Box\phi,\phi)=XQ^{2}(\phi)g(XQ^{2}(\phi)e^{\psi})-[aX^{2}Q^{5}(\phi)e^{2\psi}+rXQ^{3}(\phi)e^{\psi}]\Box\phi.\label{lagr-scal}
\end{equation}
Now, in order to ease the comparison with the literature, let us make
the following field redefinitions. First of all take $\psi\to\lambda\psi$.
Then 
\begin{equation}
\psi(\phi)=\int_{\phi}duQ(u).\label{psi-def2}
\end{equation}
Now consider $\phi\to\psi(\phi)$, with $\psi(\phi)$ given by Eq.
(\ref{psi-def2}). This implies $X\to X_{\psi}=XQ^{2}(\phi)$ and
$Q\Box\phi\to\Box\psi+2\frac{d\log Q}{d\psi}X_{\psi}$. Then Eq. (\ref{lagr-scal})
turns into 
\begin{equation}
\mathcal{L}(X_{\psi},\Box\psi,\psi)=X_{\psi}g(X_{\psi}e^{\lambda\psi})-[aX_{\psi}^{2}e^{2\lambda\psi}+rX_{\psi}e^{\lambda\psi}](\Box\psi+2\frac{d\log Q}{d\psi}X_{\psi}).\label{lagr-scal3}
\end{equation}
With these redefinitions the expression for the coupling, Eq. (\ref{Qdef})
becomes 
\begin{equation}
1=-\frac{1}{\rho_{m}\sqrt{-g}}\frac{\delta S_{m}}{\delta\psi},
\end{equation}
which would lead to a constant coupling when expressed in terms of
$\psi$. However, the influence of the coupling is explicitly present
in the form of the Lagrangian due to the presence of the term depending
on $\frac{d\log Q}{d\psi}$. This singular character appeared due
to the presence of $\Box\phi$ in the Lagrangian, and is not present
in the part of the Lagrangian depending of $K(\phi,X)$, as shown
in Ref. \cite{Amendola:2006qi}. From here on, however, we specialize
to the case of constant coupling. For constant coupling $Q$ and after
redefining $\psi$ as ${Q}\phi$, we can rewrite Eq. (\ref{lagr-scal3})
as 
\begin{equation}
\mathcal{L}(X,\Box\phi,\phi)=Xg(Y)-(aY^{2}+rY)\Box\phi.\label{LagrY}
\end{equation}
where 
\begin{equation}
Y=Xe^{\lambda\phi}
\end{equation}
and ${Q}$ is included in a redefinition of the $\lambda$ of Eq.
(\ref{lambda}) : 
\begin{equation}
\lambda={Q}\biggl(\frac{1+w_{eff}}{\Omega_{\phi}(w_{m}-w_{\phi})}\biggr).\label{lambda1}
\end{equation}
In the case of pressureless matter $w_{m}=0$ and $w_{eff}=\Omega_{\phi}w_{\phi}$
so that we obtain 
\begin{equation}
w_{eff}=-\frac{Q}{\lambda+Q}\label{eq:conj}
\end{equation}
This effective equation of state characterizes the scaling solutions.
Since this relation does not depend on the form of the Lagrangian
(just as Eq. \ref{lambda}) but rather on the solution by separation
of variables, we conjecture that adding new independent terms to the
Lagrangian will not modify it. In other words, we expect to see the
same relation $w_{eff}(\lambda,Q)$ for the entire Horndeski Lagrangian,
provided there exist non trivial solutions. 

Since the last equation is invariant under a simultaneous change of
sign of $Q$ and $\lambda$, from here on we consider $\lambda>0$
\cite{Amendola:2006qi}. Thus we have arrived, for constant coupling
$Q$, at the general form of the Lagrangian of type (\ref{action})
that allows for scaling solutions. Then, with $\mathcal{L}=K(\phi,X)-G_{3}(\phi,X)\Box\phi$,
and comparing with Eq. (\ref{LagrY}) we get 
\begin{equation}
K(\phi,X)=Xg(Y)\label{K2}
\end{equation}
and 
\begin{equation}
G_{3}(\phi,X)=aY^{2}+rY.\label{G2}
\end{equation}
We will need the following expressions: 
\begin{eqnarray}
K_{X} & = & g+g_{1}\\
G_{3,\phi} & = & 2a\lambda Y^{2}+\lambda rY\\
G_{3,X} & = & 2a\frac{Y^{2}}{X}+r\frac{Y}{X},
\end{eqnarray}
where $g_{1}$ is defined as 
\begin{equation}
g_{1}=Y\frac{dg}{dY}.\label{g1}
\end{equation}


\section{Phase-space equations}

Now in order to study the general behavior of the solutions we consider
the Lagrangian given by Eq. (\ref{lagr-scal}) in the presence of
pressureless dust. In this case, Eq. (\ref{lambda1}) can be written
as 
\begin{equation}
\lambda=\biggl(-\frac{1}{w_{\phi}\Omega_{\phi}}-1\biggr){Q}
\end{equation}
and we see that $\lambda$ is a constant for the scaling solutions
we are looking for (where both $w_{\phi}$ and $\Omega_{\phi}$ are
constants). We now define the new variables 
\begin{align}
x & =\frac{\dot{\phi}}{\sqrt{6}H}.\label{xdef}\\
y & =\frac{e^{-\lambda\phi/2}}{\sqrt{3}H}
\end{align}
and 
\begin{eqnarray}
\zeta & = & -2\lambda(2aY^{2}+rY)\label{zeta}\\
\zeta_{1} & = & 3\dot{\phi}H\biggl(2a\frac{Y^{2}}{X}+r\frac{Y}{X}\biggr)\label{zeta1}\\
 & = & \frac{\sqrt{6}}{x}(2aY^{2}+rY)=-\frac{\sqrt{6}}{2\lambda}\frac{\zeta}{x}\nonumber \\
\zeta_{2} & = & \ddot{\phi}\biggl(2a\frac{Y^{2}}{X}+r\frac{Y}{X}\biggr).\label{zeta2}
\end{eqnarray}
Then from Eq. (\ref{rhophi2}) and Friedman equations we find 
\begin{equation}
\frac{dy}{dN}=\frac{y}{2}[3-\sqrt{6}\lambda x+3x^{2}(g+\zeta-2\zeta_{2})]\label{dydNz0}
\end{equation}
and 
\begin{eqnarray}
\frac{dx}{dN} & = & \frac{3}{2}x\biggl[(1+A\zeta_{1})[1+(g+\zeta-2\zeta_{2})x^{2}]-2A(g+g_{1}+\zeta+\zeta_{1}-\zeta_{2})\biggr]+\nonumber \\
 &  & +\frac{\sqrt{6}}{2}[A(Q+\lambda)(g+2g_{1}+\zeta+2\zeta_{1})x^{2}-A\lambda\zeta_{1}x^{2}-\lambda x^{2}-AQ]\label{dxdNz0}
\end{eqnarray}
where we defined 
\begin{align}
g_{2}(Y) & =Y^{2}\frac{d^{2}g}{dY^{2}}\label{dotg2}\\
A^{-1} & =g+5g_{1}+2g_{2}+3\zeta+3\zeta_{1}+2\lambda rY-8a\lambda Y^{2}\frac{\zeta_{1}}{\zeta}.\label{Aminus}
\end{align}
Some useful relations are 
\begin{align}
w_{\phi} & =\frac{g+\zeta-2\zeta_{2}}{g+2g_{1}+\zeta+2\zeta_{1}},\label{wphi}\\
w_{\phi}\Omega_{\phi} & =x^{2}(g+\zeta-2\zeta_{2}),\label{wOmega}\\
w_{\phi} & =-1+2\frac{x^{2}}{\Omega_{\phi}}(g+g_{1}+\zeta+\zeta_{1}-\zeta_{2}).\label{wand1}\\
w_{eff} & =\frac{p_{t}}{\rho_{t}}=-1-\frac{2}{3}\frac{1}{H}\frac{dH}{dN}=x^{2}(g+\zeta-2\zeta_{2})+\frac{z^{2}}{3}.\label{weff}
\end{align}


\section{Critical points}

Critical points are obtained from the conditions 
\begin{equation}
\frac{dx}{dN}=\frac{dy}{dN}=0.\label{fixed}
\end{equation}
From Eq. (\ref{dydNz0}) we have two classes of solutions: i) $y=0$
or ii) $3-\sqrt{6}\lambda x+3x^{2}(g+\zeta-2\zeta_{2})=0$. We will
discuss the first class later on. The second possibility gives 
\begin{equation}
x=\frac{\sqrt{6}}{2\lambda}(1+\omega_{\phi}\Omega_{\phi}).\label{xcrit}
\end{equation}
This and Eq. (80) gives 
\begin{equation}
(\lambda+Q)\biggl(1+w_{\phi}\Omega_{\phi}-\frac{2\lambda x}{\sqrt{6}}\biggr)=(\Omega_{\phi}-1)\biggl(-\frac{3}{x}+\sqrt{6}(\lambda+Q)\biggr).
\end{equation}
One can easily see from Eq. (\ref{xcrit}) that the left-hand side
of the former equation is identically null. From the right-hand side
we obtain the following possibilities: \emph{i}) scalar-field dominated
solution, where 
\begin{equation}
\Omega_{\phi}=1
\end{equation}
and \emph{ii}) scaling solution, where 
\begin{equation}
\Omega_{\phi}=-\frac{Q}{w_{\phi}(\lambda+Q)}.\label{Omega_scaling}
\end{equation}
Note that the two solutions obtained here coincide with those obtained
in Ref. \cite{Amendola:2006qi} for the simpler Lagrangian with $G_{3}=0$.
In the following we will consider separately the properties of these
two classes of fixed points.

We will need a useful identity for $\zeta_{2}$ valid on the critical
points. Firstly we rewrite Eq. (\ref{zeta2}) as 
\begin{equation}
\zeta_{2}=\frac{\zeta}{2}\biggl[\frac{\sqrt{6}}{2\lambda x}(1-x^{2}(g+\zeta-2\zeta_{2}))-\frac{\sqrt{6}}{3\lambda x^{2}}\frac{dx}{dN}\biggr]\label{zeta2-b}
\end{equation}
Now this gives, together with Eqs. (\ref{wOmega}) and (\ref{xcrit}),
\begin{equation}
\zeta_{2}=\zeta_{2}(Y)=\frac{\zeta}{2}=-{\lambda}(2aY^{2}+rY)\label{idzeta2}
\end{equation}
when $dx/dN=dy/dN=0$, with $y\neq0$.


\subsection{Point A: Scalar-field dominated solutions}

For $\Omega_{\phi}=1$, Eq. (\ref{xcrit}) gives 
\begin{equation}
w_{\phi}=-1+\frac{\sqrt{6}}{3}\lambda x,
\end{equation}
and from Eqs. (\ref{weff}) we get the effective equation of state
$w_{\mathrm{eff}}=w_{\phi}$. This gives that for an accelerated expansion,
where $w_{\mathrm{eff}}<-1/3$, we must have 
\begin{equation}
\lambda x<\frac{\sqrt{6}}{3}.
\end{equation}
Since, from Eqs. (\ref{zeta1}) and (\ref{idzeta2}), $\zeta_{1},\zeta_{2}$
are functions only of $x$ and $Y$, given $g(Y)$ and $g_{1}(Y)$,
in principle we can obtain $x$ and $Y$. Also, since 
\begin{equation}
Y=\frac{x^{2}}{y^{2}},\label{Xxy}
\end{equation}
after obtaining $x$ and $Y$ we can get the scalar-field dominant
fixed-points $(x,y)$. Even for the simple models of ordinary scalar
field (where $g(Y)=1-c/Y$, with $c$ constant) and dilatonic ghost
condensate \cite{bose} (where $g(Y)=-1+cY$) the expressions found
for $(x,y)$, despite explicit, are too intrincate to be useful and
we will not present them.


\subsection{Point B: Scaling solutions}

With $\Omega_{\phi}$ given by Eq. (\ref{Omega_scaling}), Eq. (\ref{xcrit})
gives 
\begin{equation}
x=\frac{\sqrt{6}}{2(\lambda+Q)},\label{xscal}
\end{equation}
and Eq. (\ref{weff}) gives 
\begin{equation}
w_{eff}=-\frac{Q}{(\lambda+Q)}.\label{weffscal}
\end{equation}
Note that $x$ and $w_{eff}$ are independent of the explicit form
of $g,g_{1},\zeta,\zeta_{1}$ and $\zeta_{2}$. The condition for
accelerated expansion, $w_{eff}<-1/3$ leads to the following possibilities:
\begin{equation}
Q>\frac{\lambda}{2}\label{scal_up}
\end{equation}
or 
\begin{equation}
Q<-\lambda\label{scal-dn}
\end{equation}
From Eq. (\ref{wand1}), after using Eqs. (\ref{xscal}) and (\ref{weffscal}),
we have 
\begin{equation}
\Omega_{\phi}=\frac{Q(Q+\lambda)+3[g+g_{1}+(\lambda+2Q)(2aY^{2}+rY)]}{(\lambda+Q)^{2}}.\label{Omega-scal}
\end{equation}
The condition $dy/dN=0$ with $y\neq0$ in Eq. (\ref{dydNz0}) gives,
after using Eq. (\ref{xscal}) and Eq. (\ref{idzeta2}) 
\begin{equation}
g=-\frac{2}{3}Q(Q+\lambda).\label{gzeta-scal}
\end{equation}
For a given model, once $g(Y)$ is specified, we can solve Eq. (\ref{gzeta-scal})
to find $Y$. The value of $y$ is then obtained as $y=|x|/\sqrt{Y}$,
with $x$ given by Eq. (\ref{xscal}). The values of $\Omega_{\phi}$
and $w_{\phi}$ are then obtained after using Eqs. (\ref{Omega-scal})
and (\ref{Omega_scaling}). It is remarkable that Eqs. (\ref{xscal})
and (\ref{gzeta-scal}) are exactly the same obtained in Ref. \cite{st2006}
for $w_{m}=0$ and $G_{3}=0$. In this way we show that the scaling
solutions we found are not able to distinguish between the presence
of a term depending linearly on $\Box\phi$ in the Lagrangian. For
example, for the dilatonic ghost condensate with $g(Y)=-1+cY$, we
have 
\begin{equation}
Y=\frac{1}{3c}[3+2Q(Q+\lambda)]
\end{equation}
and 
\begin{equation}
y=\biggl[\frac{9c}{2(Q+\lambda)^{2}[3+2Q(Q+\lambda)]}\biggr]^{1/2}.
\end{equation}
For an ordinary scalar field with $g(Y)=1-c/Y$ we have 
\begin{equation}
Y=\frac{3c}{2Q(Q+\lambda)+3}
\end{equation}
and 
\begin{equation}
y=\biggl[\frac{2Q(Q+\lambda)+3}{2c(Q+\lambda)}\biggr]^{1/2},
\end{equation}
which coincides with the result obtained in Ref. \cite{st2006}.


\subsection{Points $C$ and $D$, for $y=0$}

When $y\to0$ we have $Y\to\infty$, and the contribution from $G_{3}$
to the Lagrangian is singular unless $a=r=0$, which recovers the
known results from the literature \cite{Amendola:2006qi}, which we
review here for the sake of completeness. One can expand $g$ in positive
integer powers of $Y$ 
\begin{equation}
g=c_{0}+\sum_{n=1}^{\infty}c_{n}Y^{-n}=c_{0}+\sum_{n=1}^{\infty}c_{n}\biggl(\frac{y^{2}}{x^{2}}\biggr)^{n},
\end{equation}
which gives, for $y=0$, to $g=c_{0}$ and $g_{1}=g_{2}=0$. Also
we have in this case $\zeta=\zeta_{1}=\zeta_{2}=0$. Condition $dx/dN=0$
then gives 
\begin{equation}
\frac{1}{2}(3c_{0}x+\sqrt{6}Q)\biggl(x^{2}-\frac{1}{c_{0}}\biggr)=0.
\end{equation}
We have the following possibilities: \emph{i}) point $C$, called
$\phi$-matter-dominated era ($\phi$MDE), where (see Ref. \cite{amendola2000})
\begin{equation}
(x,y)=\biggl(-\frac{\sqrt{6}Q}{3c_{0}},0\biggr),
\end{equation}
which leads to 
\begin{eqnarray}
\Omega_{\phi} & = & \frac{2Q^{2}}{3c_{0}},\\
w_{\phi} & = & 1,\\
w_{\mathrm{eff}} & = & \frac{2Q^{2}}{3c_{0}}.
\end{eqnarray}
\emph{ ii}) point $D$, called pure kinetic solutions, where (see
Ref. \cite{Amendola:2006qi}) 
\begin{equation}
(x,y)=\biggl(\pm\frac{1}{\sqrt{c}_{0}},0\biggr),
\end{equation}
which leads to 
\begin{eqnarray}
\Omega_{\phi} & = & 1,\\
w_{\phi} & = & w_{\mathrm{eff}}=1.
\end{eqnarray}
\begin{table}[ht]
\caption{ Critical Points for Horndeski Lagrangian}

\centering %
\begin{tabular}{|c|c|c|c|c|}
\hline 
Point  & $x$  & $y$  & $\Omega_{\phi}$  & $w_{eff}$ \tabularnewline
\hline 
A ($\phi$-dominated solutions)  & $x_{A}$  & ${(\frac{{x_{A}}^{2}}{Y_{A}})}^{1/2}$  & 1  & $-1+\frac{\sqrt{6}}{3}\lambda x$ \tabularnewline
\hline 
B (scaling solutions)  & $\frac{\sqrt{6}}{2(\lambda+Q)}$  & ${(\frac{{x_{B}}^{2}}{Y_{B}})}^{1/2}$  & $\frac{Q(Q+\lambda)+3[g+g_{1}+(\lambda+2Q)(2aY^{2}+rY)]}{(\lambda+Q)^{2}}$  & $-\frac{Q}{(\lambda+Q)}$ \tabularnewline
\hline 
C (see Refs. \cite{Amendola:2006qi}, \cite{amendola2000})  & $-\frac{\sqrt{6}Q}{3c_{0}}$  & $0$  & $\frac{2Q^{2}}{3c_{0}}$  & $\frac{2Q^{2}}{3c_{0}}$ \tabularnewline
\hline 
D (see Ref. \cite{Amendola:2006qi})  & $\pm\frac{1}{\sqrt{c_{0}}}$  & 0  & $1$  & 1 \tabularnewline
\hline 
\end{tabular}\label{table:nonlin} 
\end{table}

The Table I presents the main results for fixed points from this work
and from Ref. \cite{Amendola:2006qi} to ease the comparison.


\section{stability analysis for $y_{c}\neq0$}

Here we analyze the stability of the fixed points $A$ and $B$ obtained
in the former section. We consider small perturbations around the
critical point $(x_{c},y_{c})$ as 
\begin{eqnarray}
x & = & x_{c}+\delta x,\\
y & = & y_{c}+\delta y,\\
Y & = & Y_{c}+\delta Y.
\end{eqnarray}
We expand the function $g(Y)$ as 
\begin{equation}
g(Y)=g_{c}+g_{c}(Y-Y_{c})+\frac{g_{c}''}{2}(Y-Y_{c})^{2}+...\,\,,\label{gYc}
\end{equation}
where $g_{c}=g(Y_{c})$. Defining $\delta Y=Y-Y_{c}$, from Eq. (\ref{Xxy})
we have 
\begin{equation}
\delta Y=2\frac{Y_{c}}{x_{c}}\delta x-2\frac{Y_{c}}{y_{c}}\delta y.
\end{equation}
Then finally we obtain the following perturbation equations 
\begin{equation}
\frac{d}{dN}\biggl(\frac{\delta x}{\delta y}\biggr)=\mathcal{M}\biggl(\frac{\delta x}{\delta y}\biggr).
\end{equation}
where 
\begin{equation}
\mathcal{M}=\biggl(\begin{array}{cc}
a_{11} & a_{12}\\
a_{21} & a_{22}
\end{array}\biggr)
\end{equation}
and (the subscript $c$ means evaluated at the critical point) 
\begin{eqnarray}
a_{11} & = & \biggl[1-{\zeta_{1}}_{c}\biggl(A_{c}-\frac{{x_{c}}^{2}}{1-{\zeta_{1}}_{c}{x_{c}}^{2}}\biggr)\biggr]^{-1}\biggl[{a_{11}}\biggr|_{{G_{3}=0}}\nonumber \\
 &  & +3{\zeta_{1}}_{c}\biggl[A_{c}\biggl(2-\frac{\sqrt{6}}{2}\lambda x+g{x_{c}}^{2}\biggr)-\frac{{x_{c}}^{2}}{1-{\zeta_{1}}_{c}{x_{c}}^{2}}\biggl(1-\frac{\sqrt{6}}{6}\lambda x-g_{1}{x_{c}}^{2}\biggr)\biggr]\biggr]\\
a_{12} & = & \biggl[1-{\zeta_{1}}_{c}\biggl(A_{c}-\frac{{x_{c}}^{2}}{1-{\zeta_{1}}_{c}{x_{c}}^{2}}\biggr)\biggr]^{-1}\biggl[{a_{12}}\biggr|_{{G_{3}=0}}+\frac{{\zeta_{1}}_{c}{x_{c}}^{2}}{y}\biggl(-\frac{\sqrt{6}}{2}A_{c}(2Q+\lambda)-3\frac{{g_{1}}_{c}{x_{c}}^{3}}{1-{\zeta_{1}}_{c}{x_{c}}^{2}}\biggr)\biggr]\\
a_{21} & = & {a_{21}}\biggr|_{{G_{3}=0}}-3{x_{c}}^{2}y_{c}(1-{\zeta_{1}}_{c}{x_{c}}^{2})^{-1}\biggl(\frac{{\zeta_{1}}_{c}}{x_{c}}+\frac{\zeta_{c}}{2x_{c}}-{\zeta_{1}}_{c}{g_{1}}_{c}{x_{c}}+\frac{{\zeta_{1}}_{c}}{3x_{c}}a_{11}\biggr)\\
a_{22} & = & {a_{22}}\biggr|_{{G_{3}=0}}-3{x_{c}}^{2}y_{c}(1-{\zeta_{1}}_{c}{x_{c}}^{2})^{-1}\biggl({\zeta_{1}}_{c}{g_{1}}_{c}\frac{{x_{c}}^{2}}{y_{c}}+\frac{{\zeta_{1}}_{c}}{3x_{c}}a_{12}\biggr).
\end{eqnarray}
with 
\begin{eqnarray}
a_{11}\biggr|_{{G_{3}=0}} & = & -3+\frac{\sqrt{6}}{2}(2Q+\lambda)x_{c}+3x^{2}(g_{c}+{g_{1}}_{c})\\
a_{12}\biggr|_{{G_{3}=0}} & = & y\biggl(-3x_{c}{g_{1}}_{c}Y_{c}+3\frac{x_{c}}{y_{c}^{2}}-\sqrt{6}(Q+\lambda)Y_{c}+\sqrt{6}A_{c}\frac{(Q+\lambda)\Omega_{\phi}+Q}{2y_{c}^{2}}\biggr)\\
a_{21}\biggr|_{{G_{3}=0}} & = & \frac{y_{c}}{2}[-\sqrt{6}\lambda+6x(g_{c}+{g_{1}}_{c})]\\
a_{22}\biggr|_{{G_{3}=0}} & = & -3{x_{c}}^{2}{g_{1}}_{c}.
\end{eqnarray}
Note that for $a=r=0$ we recover the results from \cite{st2006}
for $w_{m}=0$. The eigenvalues of $\mathcal{M}$ are 
\begin{equation}
\mu_{\pm}=\xi_{1}[1\pm\sqrt{1-\xi_{2}}],\label{mupm}
\end{equation}
with 
\begin{align}
\xi_{1} & =\frac{a_{11}+a_{22}}{2}\\
\xi_{2} & =\frac{4(a_{11}a_{22}-a_{12}a_{21})}{(a_{11}+a_{22})^{2}}.
\end{align}
Stability is verified provided the conditions $\xi_{1}>0$ and $\xi_{2}>0$
are satisfied. In the following we will consider separately the stability
conditions for the two classes of fixed points.


\subsection{$\phi$-dominated solutions}

For this case we have the following eingenvalues of the matrix $\mathcal{M}$:
\begin{align}
\mu_{+} & =\mu_{+}\biggr|_{{G_{3}=0}}-3+\sqrt{6}(Q+\lambda)x,\\
\mu_{-} & =\mu_{-}\biggr|_{{G_{3}=0}}-3+\frac{\sqrt{6}}{2}\lambda x.
\end{align}
This means that the $\phi-$dominated solutions obey the same stability
conditions found in Ref. \cite{st2006}. The fixed point A is stable
if $\mu_{+}<0$ and $\mu_{-}<0$. This occurs for the following conditions
\cite{st2006}: 
\begin{eqnarray}
x<\frac{\sqrt{6}}{2(Q+\lambda)}\, &  & if\,\, Q>-\frac{\lambda}{2}\\
x<\frac{\sqrt{6}}{\lambda}\, &  & if\,\,\ -\lambda<Q<-\frac{\lambda}{2}\\
\frac{\sqrt{6}}{2(Q+\lambda)}<x<\frac{\sqrt{6}}{\lambda}\, &  & if\,\,\ Q<-\lambda.
\end{eqnarray}


\subsection{Scaling solutions}

For this case we have 
\begin{equation}
\xi_{1}=-\frac{3(\lambda+2Q)}{4(\lambda+Q)}={\xi_{1}}\biggr|_{{G_{3}=0}}
\end{equation}
and 
\begin{equation}
\xi_{2}=\frac{8}{3}(1-\Omega_{\phi})\frac{(\lambda+Q)^{3}}{(\lambda+2Q)^{2}}A_{c}[\Omega_{\phi}(\lambda+Q)+Q-(\lambda+2Q){\zeta_{1}}_{c}{x_{c}}^{2}]{(1-{\zeta_{1}}_{c}{x_{c}}^{2})(1-{\zeta_{1}}_{c}A_{c})}\mathcal{F}^{2},
\end{equation}
where 
\begin{equation}
\mathcal{F}=\frac{1}{1-{\zeta_{1}}_{c}A_{c}(1-{\zeta_{1}}_{c}{x_{c}}^{2})}.
\end{equation}
The necessary condition for fixed points for scalar solutions to be
stable is $\xi_{1}<0$ and $\xi_{2}>0$. The condition $\xi_{1}<0$
gives $Q>-\lambda/2$ or $Q<-\lambda$. This means that when the more
restrictive inequalities (\ref{scal_up}) and (\ref{scal-dn}) for
an accelerated universe are satisfied, we have $\xi_{1}<0$.

Now we analyze the condition $\xi_{2}>0$, or 
\begin{equation}
(1-\Omega_{\phi})\frac{(\lambda+Q)^{3}}{(\lambda+2Q)^{2}}A_{c}[\Omega_{\phi}(\lambda+Q)+Q-(\lambda+2Q){\zeta_{1}}_{c}{x_{c}}^{2}]{(1-{\zeta_{1}}_{c}{x_{c}}^{2})(1-{\zeta_{1}}_{c}A_{c})}>0.\label{xi2_ineq}
\end{equation}
As a guide we consider the limit $G_{3}\to0$ ($\zeta_{1}\to0$).
This means to impose the conditions 
\begin{equation}
A_{c}>0,\label{Aplus}
\end{equation}
to avoid ultraviolet instabilities \cite{pt}, and also 
\begin{equation}
1-{\zeta_{1}}_{c}{x_{c}}^{2}>0,\label{zeta1x}
\end{equation}
and 
\begin{equation}
1-{\zeta_{1}}_{c}A_{c}>0.\label{zeta1A}
\end{equation}
For an accelerated universe, conditions given by Eqs. (\ref{Aplus}),
(\ref{zeta1x}) and (\ref{zeta1A}) lead to restrictions for the coupling
Q and for the coefficients $a,r$. We have the following possibilities: 
\begin{itemize}
\item For $2aY^{2}+rY<0$:

i) $Q>\lambda/2$ or \\
 ii) $Q<-\lambda+3(2aY^{2}+rY)$ and $Q<-\lambda+1/[3A_{c}(2aY^{2}+rY)]$

\item For $2aY^{2}+rY>0$:

i) $Q<-\lambda$ or \\
 ii) $Q>\lambda/2$ and $0<2aY^{2}+rY<\lambda/2$ and $0<2aY^{2}+rY<1/(3A_{c}\lambda)$
or\\
 iii) $Q>-\lambda+3(2aY^{2}+rY)$ and $2aY^{2}+rY>\lambda/2$ and
$0<2aY^{2}+rY<1/(3A_{c}\lambda)$;

\end{itemize}
Eq. (\ref{xi2_ineq}) then gives 
\begin{equation}
(1-\Omega_{\phi})(\lambda+Q)^{3}[\Omega_{\phi}(\lambda+Q)+Q-(\lambda+2Q){\zeta_{1}}_{c}{x_{c}}^{2}]>0.\label{zeta2-ineq2}
\end{equation}
We impose $\Omega_{\phi}<1$ (following \cite{st2006}). Eq. (\ref{zeta2-ineq2})
leads also to the following possibilities: \emph{i}) $Q<-\lambda$,
which is Eq. (\ref{scal-dn}) for an accelerated universe; \emph{ii})
$\Omega_{\phi}(\lambda+Q)+Q-(\lambda+2Q){\zeta_{1}}_{c}{x_{c}}^{2}>0$,
which gives 
\begin{equation}
\biggl[-\frac{Q}{Q+\lambda}+3\frac{2Q+\lambda}{(Q+\lambda)^{2}}(2aY^{2}+rY)\biggr]<\Omega_{\phi}.\label{ineqzeta}
\end{equation}
The former equation leads to $3(g+g_{1})>-2Q(Q+\lambda)$, which,
for $G_{3}=0$, is automatically satisfied for a nonphantom field
where $g+g_{1}>0$ \cite{Amendola:2006qi}.

The condition $\Omega_{\phi}<1$ give the more stringent condition
for fixed points with scaling solutions to be stable, namely, 
\begin{equation}
g+g_{1}<\frac{\lambda(Q+\lambda)}{3}-(2Q+\lambda)(2aY^{2}+rY).
\end{equation}


\section{stability analysis for $y_{c}=0$}

In this section, for completeness, we review the stability analysis
of the fixed points $C$ and $D$ obtained in the literature\cite{Amendola:2006qi}.
We remember that we are considering here $a=r=0$. After small perturbations
around the critical point $(x_{c},y_{c})$ we have 
\begin{align}
\frac{d(\delta x)}{dN} & =\biggl(-\frac{3}{2}+9c_{0}x^{2}+\sqrt{6}Qx\biggr)\delta x\\
\frac{d(\delta y)}{dN} & =\frac{3}{2}\biggl(1+c_{0}x_{c}^{2}-\frac{\sqrt{6}}{3}\lambda x\biggr)\delta y.
\end{align}
This means that in the matrix $\mathcal{M}$ of perturbations we have
$a_{12}=a_{21}=0$. Then, from Eq. (\ref{mupm}) the eigenvalues of
$\mathcal{M}$ are 
\begin{align}
\mu_{+} & =a_{11}=-\frac{3}{2}+9c_{0}x^{2}+\sqrt{6}Qx\\
\mu_{-} & =a_{22}=\frac{3}{2}(1+c_{0}x_{c}^{2}-\frac{\sqrt{6}}{3}\lambda x).
\end{align}
The main results for points C and D are the following: i) For point
C, condition $\Omega_{\phi}<1$ gives $|Q|<\sqrt{3c_{0}/2}$ for $c_{0}>0$.
This leads, in case of accelerated expansion ($w_{eff}<-1/3$), to
$\mu_{+}<0$ and $\mu_{-}>0$, a condition for saddle point. For $c_{0}<0$
we have $\mu_{+}<0$; this with the condition $Q(Q+\lambda)>3|C_{0}|/2$
leads to $\mu_{-}<0$, resulting the fixed point as a stable node.
ii) For point D (which exists only for $c_{0}>0$), we have the following
possibilities: a) if $Q>0$ we have at least one of $\mu_{+}$ and
$\mu_{-}$ to be positive. This leads to unstable nodes or saddle
points depending on the values of $\lambda,Q$. b) If $Q<-\sqrt{3c_{0}/2}$
and $\lambda>\sqrt{6c_{0}}$ the point $x=1/\sqrt{c_{0}}$ is stable.
c) if $Q<0$ the point $x=-1/\sqrt{c_{0}}$ is unstable. The Table
II presents the main results for stability analysis of fixed points
from this work and from Ref. \cite{Amendola:2006qi} to ease the comparison.
\begin{table}[ht]
\caption{ Stability Analysis of Critical Points for Horndeski Lagrangian}

\centering %
\begin{tabular}{|c|c|c|}
\hline 
Point  & Stability  & Existence \tabularnewline
\hline 
A ($\phi$-dominated)  & Stable node for  & \tabularnewline
 & $x<\frac{\sqrt{6}}{2(Q+\lambda)}$ if $Q>-\frac{\lambda}{2}$  & $y\neq0$ and $\Omega_{\phi}=1$ \tabularnewline
 & $x<\frac{\sqrt{6}}{\lambda}$ if $-\lambda<Q<-\frac{\lambda}{2}$  & \tabularnewline
 & $\frac{\sqrt{6}}{2(Q+\lambda)}<x<\frac{\sqrt{6}}{\lambda}$ if $Q<-\lambda$  & \tabularnewline
\hline 
B (scaling)  & Stable node for  & \tabularnewline
 & $-\frac{Q}{Q+\lambda}+3\frac{2Q+\lambda}{(Q+\lambda)^{2}}(2aY^{2}+rY)<\Omega_{\phi}<1$  & $g+g_{1}<\frac{1}{3}{\lambda(Q+\lambda)}-(2Q+\lambda)(2aY^{2}+rY)$ \tabularnewline
 & with restrictions on $Q,a,r$ that  & \tabularnewline
 & follows from Eqs. (\ref{Aplus}), (\ref{zeta1x}) and (\ref{zeta1A})  & \tabularnewline
\hline 
C (from Refs. \cite{Amendola:2006qi}, \cite{amendola2000})  & Saddle point for $c_{0}>0$  & $a=r=0$ because $G_{3}\neq0$ is singular \tabularnewline
 & Stable node for $c_{0}<0$ and $Q(Q+\lambda)<3|c_{0}|/2$  & $|Q|<\sqrt{3c_{0}/2}$ or $c_{0}<0$ \tabularnewline
\hline 
D (see Ref. \cite{Amendola:2006qi})  & Unstable node or saddle point for $Q>0$  & $a=r=0$ because $G_{3}\neq0$ is singular \tabularnewline
$x=\frac{1}{\sqrt{c_{0}}}$  & Stable node for $Q<-\sqrt{3c_{0}/2}0$ and $\lambda>\sqrt{6c_{0}}$  & $c_{0}>0$ \tabularnewline
D (see Ref. \cite{Amendola:2006qi})  & Unstable node or saddle point for $Q>0$  & $a=r=0$ because $G_{3}\neq0$ is singular \tabularnewline
$x=-\frac{1}{\sqrt{c_{0}}}$  & Unstable node for $Q<0$  & $c_{0}>0$ \tabularnewline
 &  & \tabularnewline
\hline 
\end{tabular}\label{table:nonlin2} 
\end{table}

\section{The possibility of two scaling regimes}

Now we search for the possibility of two successive cosmologically
viable scaling regimes: one dominated by matter and other dominated
by dark energy. Such a transition would allow for a standard matter
era before the onset of acceleration, which in turn would help obtaining
a standard growth of perturbation in this class of models. Ref. \cite{Amendola:2006qi}
investigated this possibility for the case $a=r=0$. In the case of
$g(Y)$ approximated by a polynomial with positive and negative powers
of $Y$, Ref. \cite{Amendola:2006qi} showed that this is not possible.
Here we consider again this possibility in the extended context of
Horndeski Lagrangian (\ref{action-1}). The existence of a matter-dominated
phase is characterized by 
\begin{equation}
\Omega_{\phi}=x^{2}(g+2g_{1}+\zeta+2\zeta_{1})=0
\end{equation}
and 
\begin{equation}
w_{\mathrm{eff}}=x^{2}(g+\zeta+2\zeta_{2})=x^{2}g=0,
\end{equation}
which gives two possibilities: i) $x=0$ or ii) $g=0$ and $2g_{1}+\zeta+2\zeta_{1}=0$.
Eqs. (\ref{dydNz0}) and (\ref{dxdNz0}) gives, for critical points
with $\Omega_{\phi}=0$: 
\begin{equation}
\sqrt{6}x_{c}(g_{c}+{g_{1}}_{c}+\zeta_{c}+{\zeta_{1}}_{c}-{\zeta_{2}}_{c})=-Q.
\end{equation}
If we consider $g(Y)$ described in terms of a series of positive
integer powers of $Y$, namely 
\begin{equation}
g=c_{0}+\sum_{n=1}^{\infty}c_{n}Y^{n},\label{gpos}
\end{equation}
then $y=0$ is forbidden, which excludes points $C$ and $D$. Each
one from the two possibilities i) and ii) leads to $Q=0$. This shows
that the choice given by Eq. (\ref{gpos}) does not satisfy conditions
given by Eqs. (\ref{scal_up}) and (\ref{scal-dn}) for an accelerated
universe.

Now we consider instead a function $g(Y)$ described in terms of a
series of negative integer powers of $Y$, namely 
\begin{equation}
g=c_{0}+\sum_{n=1}^{\infty}c_{n}Y^{-n}.\label{gneg}
\end{equation}
For $a=r=0$ this allows for $y=0$ and points C and D. However, Ref.
\cite{Amendola:2006qi} showed that in this case the decelerated phase
for point C cannot be followed by the accelerated phase given by point
B without crossing $x=0$, which means a singularity for $g$. For
the general case $a\neq0$ or $b\neq0$ the points $C$ and $D$ (where
$y=0$) are excluded due to the presence of positive polinomial powers
of $Y$ in $G_{3}$. Then the only possible critical points with acceleration
are points $A$ and $B$, and $A$ is not cosmologically viable. This
shows that the inclusion of the term $G_{3}\Box\phi$ from Horndeski
Lagrangian does not changes the conclusions of Ref. \cite{Amendola:2006qi}
concerning to the absence of a sequence of scaling regimes.

\section{Scaling solution in Einstein and Jordan Frames}

\label{sec:So-far-we}So far we have been working in the so-called
Einstein frame, where the gravitational sector is the standard Einstein-Hilbert
term. If one performs a conformal transformation 
\begin{equation}
\hat{g}_{\mu\nu}=e^{2\omega}g_{\mu\nu}
\end{equation}
with 
\begin{equation}
\omega=-Q\phi
\end{equation}
the matter Lagrangian can be decoupled, while a Brans-Dicke term $e^{2Q\phi}R$
appears in the Jordan-frame Lagrangian. As shown in e.g. \cite{amendola1999,APT},
the conformal transformation induces the following transformation
on the quantities that characterize the FLRW metric: 
\begin{equation}
\hat{\rho}_{m}=e^{4\omega}\rho_{m}\,,\;\hat{p}_{m}=e^{4\omega}p_{m}\,,\; d\hat{t}=e^{-\omega}dt\,,\;\hat{a}=e^{-\omega}a\,\;
\end{equation}
It is then easy to derive the following transformations: 
\begin{align}
\hat{H} & =e^{\omega}H(1-x)\\
\hat{\dot{H}} & =e^{2\omega}H^{2}[(1-x)(x+\frac{\dot{H}}{H^{2}})-\frac{\dot{x}}{H}]
\end{align}
where $x=\frac{\dot{\phi}}{\sqrt{6}H}=-\frac{\dot{\omega}}{\sqrt{6}QH}.$
This allows us to find the general relation between the equation of
state of the Einstein and Jordan frames: employing the definition
in Eq. (\ref{dotHH2}) we find 
\begin{equation}
\hat{w}_{eff}=-1-\frac{2}{3(1-x)}\left(x+\frac{\dot{H}}{H^{2}}-\frac{\dot{x}}{H(1-x)}\right)
\end{equation}
where 
\begin{equation}
\frac{\dot{H}}{H^{2}}=-\frac{3}{2}(1+w_{eff})
\end{equation}
In a scaling regime $\dot{x}=0$ and therefore 
\begin{equation}
\hat{w}_{eff}=\frac{x+3w_{eff}}{3(1-x)}
\end{equation}
If the field is static, i.e. $x=0$, the conformal transformation
becomes trivial and the two equations of state coincide. Analogously,
we can see that 
\begin{equation}
\hat{\Omega}_{m}=\frac{\Omega_{m}}{(1-x)^{2}}
\end{equation}
These relations allow to transforms our results from a frame to another
and show that a scaling solution in a frame is scaling also in the
other one, although with different values of $\Omega_{m},w_{eff}$.
It is interesting to note that if the expansion is accelerated in
the Einstein frame, i.e. if $w_{eff}<-1/3$, then it is accelerated
also in the Jordan frame since $|x|\le1$ and therefore $\hat{w}_{eff}<-1/3$.

We can also notice that the frame that should be compared to observations
is the one in which baryons follow geodesics, since observations are
obviously made assuming that the masses of particles remain constant
(for instance, the spectroscopic lines that determine the redshift
scale linearly with the electron mass, see the discussion in \cite{wett2013}).
If the coupling is universal and no screening mechanism is at work,
then the ``observed'' frame is the Jordan one. If however baryons
are decoupled or the local coupling in dense objects (e.g. stars)
is screened then the observed frame might as well be the Einstein
frame.

\section{Conclusions}

In this work we have investigated scaling solutions for a KGB Lagrangian,
i.e. a subclass of the  Horndeski Lagrangian with a linear dependence
in $\Box\phi$ and coupling $Q=-1/({\rho_{m}\sqrt{-g}})\frac{\delta S_{m}}{\delta\phi}$
between pressureless matter and the field that carries dark energy.
We have found a master equation for the pressure $p(X,\Box\phi,\phi)$
that satisfy the scaling condition $\Omega_{\phi}/\Omega_{m}=$ constant.
We also assumed that for asymptotic scaling solutions the equation
of state parameter $w_{\phi}$ is a constant. After a convenient Ansatz
and linearity considerations, and assuming a constant universal coupling,
we applied usual separation of variables and we found a general solution
for the Lagrangian density given by 
\begin{equation}
\mathcal{L}(X,\Box\phi,\phi)=Xg(Y)-(aY^{2}+rY)\Box\phi
\end{equation}
with $Y=Xe^{\lambda\phi}$ and $a,r$ two arbitrary constants. After
a rescaling of the field, the general form of the Lagrangian extends
known results from the literature. In order to study the general behavior
of the solutions we rewrote Friedman and field equations in terms
of dimensionless variables $(x,y)$. The fixed points, defined by
the conditions $dx/dN=dy/dN=0$ where obtained in the absence of radiation.
For $y\neq0$ we have found two classes of fixed points: A) scalar
field dominated solutions, where $\Omega_{\phi}=1$ and B) scaling
solutions, where $\Omega_{\phi}=-Q/{[w_{\phi}(\lambda+Q)]}$. Solutions
with $y=0$ are not possible with the extension of $\Box\phi$ in
the Lagrangian due to the presence of singularities. This means that
we must have $a=r=0$ in order to recover two other possible solutions:
C) $\phi$ MDE solutions and D) pure kinetic solutions.

We have shown that the scaling solution in this class of Lagrangians
has the same effective equation of state $w_{eff}$ (\ref{eq:conj})
of the Lagrangian without the $\Box\phi$ term, depending only on
the coupling $Q$ and on the exponent $\lambda$. We conjecture that
the same relation holds for the entire Horndeski Lagrangian. Moreover,
we extend to this Lagrangians the conclusion that a transition from
a matter epoch to a scaling epoch is not possible. If a component
of uncoupled baryons is included, then we would obtain an epoch of
baryon domination after the radiation era and before the scaling attractor,
as in \cite{Amendola:2001rc}. Whether this trajectory is a globally
acceptable cosmological solution is still to be ascertained. One must
remark that our conclusions are restricted to couplings with one scalar
field. For instance, considering couplings with a scalar and a vector
field that has a background isotropy-violating component \cite{thor},
the sequence radiation domination $\to$ anysotropic matter-domination
$\to$ isotropic scaling dark energy domination attractor can be realized
for a convenient choice of parameters.

\section{acknowledgements}

Gomes was sponsored by CNPq, National Council for Scientific and Technological
Development - Brazil. He is also grateful to ITP, Universitaet Heidelberg
for hospitality during the development of this project. L.A. acknowledges
support from DFG through the TransRegio33 ``The Dark Universe''
project.

\end{document}